\documentclass{aa}
\usepackage{natbib}
\usepackage{graphicx}
\newcommand{\rebf}[1]{#1}
\newcommand{\aabf}[1]{#1}
\begin{document}
\title{The photospheric abundances of active binaries}

\subtitle{I. Detailed analysis of HD 113816 (IS Vir) and HD 119285
  (V851 Cen)}

\author{D. Katz \inst{1,2} \and
  F. Favata \inst{1} \and
  S. Aigrain \inst{1,3} \and
  G. Micela \inst{4}
  }

\offprints{D. Katz,\\ \email{david.katz@obspm.fr}}

\institute{Astrophysics Division -- Research and Science Support
  Department of ESA, ESTEC, 
  Postbus 299, NL-2200 AG Noordwijk, The Netherlands
  \and
  Observatoire de Paris, GEPI, Place Jules Janssen,
  F-92195 Meudon, France 
  \and
  Institute of Astronomy, Madingley Road, Cambridge,
  United Kingdom
  \and
  Osservatorio Astronomico di Palermo, 
  Piazza del Parlamento 1, I-90134 Palermo, Italy 
  }

\date{Received ; accepted}

\abstract{The high-resolution optical spectra of the two X-ray active
  binaries RS CVn stars HD 113816 (IS~Vir) and HD 119285 (V851~Cen)
  are analysed and their Na, Mg, Al, Si, Ca, Sc, Ti, Co and Ni
  contents determined, in the framework of a larger program of chemical
  analysis of RS CVn \aabf{stellar} atmosphere.
  The analysis of IS~Vir and V851~Cen
  is performed with three different LTE methods. In the first one,
  abundances are derived for a large set of transitions (among which 28
  Fe~{\sc i} lines, spanning a broad interval in excitation potential and
  equivalent width, and 6 Fe~{\sc ii} transitions) using measured equivalent
  widths and Kurucz LTE model atmospheres as input for the MOOG software
  package. The input atmospheric parameters and abundances are iteratively
  modified until (i) the Fe~{\sc i} abundances exhibit no trend with
  excitation potential or equivalent width, (ii) Fe~{\sc i} and
  Fe~{\sc ii} average abundances are the same and (iii) Fe and
  Alpha elements average abundances are consistent with the input values.
  The second method follows a similar approach, but uses a restricted
  line list (without the Fe~{\sc i} ``low excitation potential''
  transitions) and relies on the $B-V$ and $V-I$ colour indices to determine
  the temperature. The third method uses the same restricted line list
  as the second method and relies on fitting the 6162 \AA\ Ca {\sc i} line wing
  profiles to derive the surface gravity. The reliability of these
  methods is investigated in the context of single line RS CVn stars. It
  is shown that the $V-I$
  photometric index gives, on a broader sample of stars, significantly
  cooler estimates of the effective temperature than the $B-V$ index.
  All   approaches give results in good agreement with each other,
  except the $V-I$ based method. The analysis of IS~Vir and V851~Cen results
  in both cases in their primaries being giant stars of near-solar
  metallicity. Their parameters as derived with the first method
  are respectively $T_{\rm eff} = 4720$ K, $\log g = 2.65$,
  ${\rm [Fe/H]} = +0.04$ and $T_{\rm eff} = 4700$ K, $\log g =
  3.0$ and ${\rm [Fe/H]} = -0.13$. In the case of V851~Cen the
  derived iron content is significantly higher than a previous
  determination in the literature. Both stars exhibit relative
  overabundances of several elements (e.g. Ca) with respect to the
  solar mix.
  }


\maketitle
%
%

\section{Introduction} \label{sec:intro}

Since the advent of high-energy astronomy, active binaries have been
studied in more and more detail thanks to their very high activity
levels, several orders of magnitude higher than e.g.\ in the Sun.  In
the Sun the modest rotation level induces a correspondingly modest
level of \rebf{magnetic} activity and sunspot cover, even at solar maximum,
a modest fraction of the solar surface.
In active binaries on the other hand the tidally locked rotation is
likely to induce significant transformations in the stellar structure,
and also manifests itself through much higher levels of magnetic activity.
Doppler imaging techniques have also produced evidence for very large
starspot groups (see e.g. \citealp{ri01} and previous papers in the series).
A good understanding of the structure and evolutionary state of active
binaries can thus help to understand coronal phenomena at their
extreme levels.

One new area of investigation which has opened up since the advent of
enhanced resolution X-ray spectroscopy is the study of the metal
abundance in coronal plasmas. Abundance values significantly lower than
solar have been derived (see e.g.\ \citealp{jo98} for a review).
In the Sun it is observed that the coronal abundances are not the same
as the photospheric ones, with elements being selectively enhanced
(\citealp{fe98}) on the basis of their first ionisation potential (FIP).
\rebf{The possibility that some elements are actually depleted is also
not ruled out (\citealp{jo98}).} Detailed individual coronal
abundances of other stars
are now becoming available, and thus a new question arises: the relationship
between photospheric and coronal abundances.

To study this issue it is\rebf{,} however\rebf{,} obviously
necessary to have a detailed knowledge of the
\emph{photospheric} abundances of the stars under investigation.
The iron photospheric abundances of most of the active stars which
are the subject of detailed investigation in e.g.\ X-rays,
are not at all well known, and the situation is worse for the other
metals. There are several reasons
for this, including (for the active binaries) the difficulty of
performing detailed spectroscopic studies of binary stars with a
significant rotation velocity, which tends to broaden and blend the
spectral lines, making the analysis difficult. Note that, in the case
of cool active stars, photometric abundance indices seem to produce
biased results, giving lower abundances than determined
spectroscopically (\citealp{fa97}) and thus cannot be used for a
detailed comparison of photospheric and coronal abundances.

\citet{ra94} have conducted a study on a
significant sample of objects (67 components in 54 systems), \rebf{using
effective temperatures derived from the $B-V$ index (which is,
however,} sensitive to both
metal abundances and interstellar reddening) and surface gravities
estimated from the spectral type and the luminosity class or when
radii and masses were known these were directly computed.  Iron and
lithium abundances were derived by comparison with a grid of synthetic
spectra covering a domain of 25 \AA\ around the 6708 \AA\ Li\,{\sc i}
doublet. The main result from the above study is that a large fraction
of its sample is composed of metal deficient stars (around 60\% of the
stars display a metallicity equal or lower than $-0.4$ dex). At the
same time a large fraction of their stars are Li rich.

\citet{ot98} studied a much more limited sample of (in general hotter)
objects, but with a much more detailed and purely spectroscopic
analysis, determining Fe, Mg and Si abundances through an equivalent
widths based analysis. Temperatures were derived fitting the wings of
the $H\alpha$ and $H\beta$ Balmer lines, and gravities were estimated
in a similar way, by comparing the wings of two strong lines to
synthetic profiles. Their study presents results for 5 systems in
common with \citet{ra94}, deriving, in 4 cases out of the 5, [Fe/H]
abundances larger by 0.1 to 0.3 dex. They also compared the
metallicities of the photosphere and corona (the latest extracted from
the literature) in II Peg and $\lambda$ And, showing that the X-ray
derived coronal abundance is lower than the photospheric value by one
order of magnitude.

We have thus started a detailed study of the photospheric abundances
in a sample of 28 RS CVn stars, aiming
at providing consistently-derived abundances of several elements in their
photosphere. They will be useful to address the issues of RS CVn
metallicity distribution, abundance pattern(s) and mixing
processes. They will also serve as a starting point
for the comparison with the coronal abundances once our sample objects will
have been observed by either \emph{Chandra} or XMM. The sample
selection as well as the observations are presented in
Sect.~\ref{sec:selec}.

RS CVn stars are \rebf{characterised by a high level of activity, which may
modify their photospheric properties.} Therefore, the
methods commonly used for the study of non active stars should be
carefully checked before being applied to RS CVn.
The aim of this paper is to compare different spectral analysis
techniques, to investigate their reliability
in the study of single line RS CVn stars. In order to do so,
two stars of the sample, the slow rotators, \rebf{single lines spectroscopic
binaries} (SB1) IS~Vir and V851~Cen,
are analysed using three different methods (Sect. \ref{sec:data}).
In addition, the consistency of the $B-V$, $V-R$ and $V-I$ indices as
temperature indicators is studied on a larger set of 8 single lined
sample stars (Sect. \ref{sub:photo}). The results of the three methods
are compared in Sect. \ref{sub:res}. The parameters and chemical
compositions of  IS~Vir and V851~Cen are discussed in Sect.~\ref{sec:dis}.
The analysis of the remaining SB1 sample stars as well as the study
of \rebf{double lines spectroscopic binaries} (SB2) will be the subject of
future papers.

\section{Sample selection and observations} \label{sec:selec}

The most extensive catalog of active binaries remains the one of
\citet{st93}, which we have thus used as parent
sample for our observing program. We have selected, from it, 28
systems (18 SB1 and 10 SB2) , bright enough ($V \la 10$) to be observed
with the instrument used (FEROS on the ESO 1.52 m telescope) at
sufficiently high $S/N$ in a reasonable time (1 hr maximum). We have
made sure that a number of systems in our sample were also present in
the work of \citet{ra94}, in order to simplify the 
comparison of the two studies. The selected stars are all systems with
moderate projected rotational velocity ($v \sin i < 30$ km s$^{-1}$),
to limit the blending of the lines.

The spectra were acquired in January 2000, with the fibre fed
cross-dispersed echelle spectrograph FEROS, mounted on the ESO 1.52 m
telescope at La Silla. The spectral range covered is 3600--9200 \AA,
with a resolution of 48\,000. The spectral reduction (i.e. bias
subtraction, flat field correction, order extraction and merging and
wavelength calibration) has been performed during the observations
with the {\it ``FEROS Data Reduction Software''} 
(URL~: http://www.ls.eso.org/lasilla/Telescopes/2p2T/E1p5M/
FEROS/offline.html) available on the
mountain. Most of the targets were observed 2 or 3 times (for a total
of 49 spectra) to detect possible modifications of line profiles and
allow a good filtering of the cosmic rays.

In this paper, 2 SB1 sample stars, IS~Vir and V851 Cen, are analysed.
Table \ref{tab:cat} summarises their visual apparent magnitudes,
number of exposures, mean resulting signal to noise ratios
(per pixel) and projected rotational velocities, $v \sin i$,
(from \citealp{st93}).

\begin{table}[h!]
\caption{Visual apparent magnitudes, number of exposures, mean resulting
signal to noise ratios (per pixel) and projected rotational velocities
\aabf{(from \citealp{st93})} of IS~Vir and V851~Cen.}
\begin{tabular}{c c c c c c} \hline
HD Num.   & Id.      & V    & Nb  & S/N & $v \sin i$    \\
          &          &      &     &     & (km s$^{-1}$) \\ \hline
HD 113816 & IS Vir   & 8.32 & 3   & 170 &  6.0            \\
HD 119285 & V851 Cen & 7.67 & 1   & 150 &  6.5          \\ \hline
\end{tabular}
\label{tab:cat}
\end{table}

\section{Data analysis} \label{sec:data}
\subsection{Methods} \label{sub:meth}

Three different approaches were used for the analysis of IS~Vir and V851~Cen.

The first one is a ``classical'' iterative LTE analysis. Measured equivalent
widths of 10 different elements (among which 28 Fe {\sc i} lines, spanning a
broad interval in excitation potential and equivalent widths, and 6
Fe~{\sc ii} transitions) were converted in abundances, using the MOOG
(\citealp{sn73}) software packages. The complete list of lines used is
given in Table \ref{tab:ato}, while their selection is discussed in Sect.
\ref{sub:lines}. MOOG uses model atmospheres and
atomic data (wavelength, excitation potential, gf values) to compute
theoretical curves of growth, from which the abundances are derived.
Kurucz LTE
plane parallel models (\citealp{ku93cd13}, see Sect. \ref{sub:model})
and Kurucz atomic data (\citealp{ku95cd23}) were used, except for gf
values which were, in majority, adjusted \rebf{by comparison with the solar
spectrum} (Sect. \ref{sub:loggf}).\\
The atmospheric parameters and abundances were obtained by iteratively
modifying the effective temperature, surface gravity, metallicity,
mean alpha element overabundance and micro-turbulence velocity of the
input model and rederiving the abundances until
(i) the Fe~{\sc i} transition abundances exhibited no
trend with excitation potential
or logarithm of reduced equivalent width\footnote{the equivalent width divided
by the wavelength of the transition}, (ii) Fe~{\sc i} and Fe~{\sc ii}
lines gave the same average abundances (ionisation equilibrium) and (iii)
the iron and alpha elements average abundances were consistent
with those of the input model atmosphere.\\
Each of \rebf{these} diagnostics is sensitive to
changes of several of the input
atmospheric parameters. \rebf{Given a set of observed equivalent widths,}
the slope of Fe~{\sc i} abundance as \aabf{a} function of
excitation potential decreases with effective temperature, is very
slightly sensitive to surface gravity and metallicity and increases
with micro-turbulence. The slope of Fe~{\sc i} abundances as \aabf{a} function
of logarithm of reduced equivalent width increases with effective temperature,
decreases with surface gravity, is very slightly sensitive to metallicity
and strongly decreases with micro-turbulence. Around the convergence
point, the mean Fe~{\sc i} abundance is \rebf{weakly} sensitive to changes
of the atmospheric parameters, while the mean Fe~{\sc ii} abundance
decreases \rebf{significantly} with effective temperature and increases
\rebf{significantly} with surface gravity.

``Low excitation potential'' neutral iron group transitions have been
reported to yield, in giant stars, lower abundances than their ``high
excitation'' counterparts (\citealp{ru80}, \citealp{dr91}).
This behaviour is usually attributed to non-LTE effects
due to the low density of the photosphere. As presented in
Sect.~\ref{sec:dis}, the first analysis of IS~Vir and V851~Cen
leads to the result that both of them are giants.
It was therefore necessary to test whether the results of the first
analysis had been affected by non-LTE effects, via the ``low excitation
potential'' transitions. Both stars were then reanalysed without
the ``low excitation potential'' transitions. Following the \citet{ru80}
study of LTE departure as \aabf{a} function of excitation potential,
all Fe\,{\sc i} lines with $\chi < 3.5$ eV were discarded from the
initial selection. This, of course, makes it impossible to rely on the
\rebf{slope of the Fe {\sc i} transition abundances as \aabf{a} function
of excitation potential
to constrain the atmospheric parameters,} because the
remaining interval, $3.6 \leq \chi \leq 5.0$ eV, is too limited.
Another ``diagnostic'' was therefore needed to avoid the LTE analysis
solution being degenerated.\\
In IS~Vir and V851~Cen second analysis, photometric colour indices were used.
The effective temperatures were derived from the $B-V$ index for IS~Vir,
and from the $B-V$
and $V-I$ indices for V851~Cen (the two colours leading
to different temperatures, V851~Cen was analysed two times, using each index
separately). Surface gravities, micro-turbulences and abundances were
estimated in the same way as in the first method, using the same set of
lines (\rebf{but excluding} the ``low excitation potential'' Fe {\sc i}
transitions).  In the case of the $B-V$/temperature transformation,
which is metallicity
sensitive, the two steps were iterated until convergence. The conversion of
photometric indices into effective temperatures, as well as the consistency
of different colour indices as RS CVn stars temperature estimator is
discussed in more details in Sect. \ref{sub:photo}.

As discussed in Sect. \ref{sub:photo}, the photometric indices present
a major drawback~: in several RS CVn stars, different colours yield
different effective temperatures. Therefore, analyses based on
photometric indices are not consistent and reliable enough to identify
any possible non-LTE effect that might affect the first method.
IS~Vir and V851~Cen were thus analysed a third time.\\
Instead of the ``low excitation potential'' Fe {\sc i} transitions or
the photometric colour indices, the third method \rebf{makes}
use of information
contained in the wings of the 6162~\AA\ Ca~{\sc i} transition.
In IS~Vir and V851~Cen, the 6162~\AA\ Ca~{\sc i} line is a ``strong'' line,
and therefore its wing profiles are strongly sensitive to surface
gravity. As the 6162~\AA\ Ca~{\sc i} wing profiles are also sensitive
to effective temperature, micro-turbulence velocity and iron
and calcium abundances, the analysis was performed in an iterative way.
The atmospheric parameters determined with the second method were
used as starting parameters. In a first step, the surface gravity was
derived by comparing the 6162~\AA\ Ca~{\sc i} line wing profiles to a
library of synthetic profiles (this step is detailed in Sect.
\ref{sub:grav}). 
\rebf{In a second step, the measured equivalent widths of the set of lines
used in the second method were converted to abundances, using the LTE
analysis program MOOG. The surface gravity of the MOOG input atmospheric
model was the one derived during the first step.}
Effective temperature, surface gravity, micro-turbulence
velocity and abundances were obtained by iteratively modifying the
input $T_{\rm eff}$, $\xi$, ${\rm [Fe/H]}$ and ${\rm [Ca/H]}$ and rerunning
the two steps until (i) the Fe~{\sc i} transitions exhibited no trend
with logarithm of reduced equivalent width, (ii) Fe~{\sc i} and
Fe~{\sc ii} lines gave the same average abundances and (iii) the iron
and calcium average abundances were consistent with the input abundances.

Another ``classical diagnostic'' is often used to derive the
effective temperature : the adjustment of the Balmer lines wing
profiles. In F- and G-type stars, the wings of the Balmer lines are
very sensitive to the effective temperature (the width of the wings
monotically increasing with $T_{\rm eff}$), while very slightly sensitive
to the surface gravity or the metallicity. As a consequence Balmer lines
are very good temperature indicators above $T_{\rm eff} \simeq 5000$ K
(\citealp{co02}),
and were effectively used by, \rebf{e.g.,} \citet{ot98} in their analysis
of hotter active binaries. Below this threshold, Balmer line wings
become less temperature sensitive and are no longer useful
as estimators. Since most of the 18 SB1 stars in our present sample are
classified early K-type stars, Balmer line profile adjustment could not
be used to derive their temperatures. However, even at higher temperature,
the presence of activity (filled in lines, or emission) makes it necessary
to be very careful in the use of Balmer lines as temperature indicators in
RS CVn stars.

\subsection{Lines selection} \label{sub:lines}

Both IS~Vir and V851~Cen are relatively cool
and therefore exhibit a lot of blended features.  A first selection of
unblended or ``slightly'' blended lines (lines that can be reliably
deblended) has therefore been performed starting from a synthetic
spectrum of parameters $T_{\rm eff} = 4500$ K, $\log g = 4.0$ and
${\mathrm {\rm [Fe/H]}} = 0.0$, covering the wavelength range
5000--8000 \AA\ (the bluer part of the spectrum suffers so heavily
from blending as to be effectively useless in the present context).
This spectrum was computed with Piskunov's {\sc synth} program
using atomic parameters from the VALD database
(\citealp{pi95}; \citealp{ku99}; 
\citealp{ku00}) and a Kurucz \aabf{model atmosphere} (\citealp{ku93cd13}).
The lines chosen in the
synthetic spectrum were then examined, one by one, in the two sample
spectra. For each star, \rebf{lines} which appeared asymmetric or showed an
unusually large width, were assumed blended with unidentified lines
and therefore discarded from the initial sample.  Lines whose profiles
overlapped \rebf{with} the position of one of the telluric lines listed in
{\it ``Rowland's table of the solar spectrum''}
(\citealp{mo66}) were also discarded. Note that the lines concerned are
not necessarily the same from star to star because of the radial
velocity differences.

Ideally, one would like to rely only on weak lines (lines on the
linear part or close to the knee of the curve of growth, i.e.\ 
approximately $W < 70$ m\AA) to perform detailed analysis.
Unfortunately, in the case of ``cool'' stars exhibiting a moderate
rotational velocity, the number of reliable and unblended lines is
rather small. 
Even for the two slow rotators analysed here, IS~Vir and V851~Cen, Na
and Mg are derived from ``strong'' lines (respectively around 100 and
150 m\AA). In order to quantify the effect of ``strong'' lines on the
results, V851~Cen has been analysed (method 3) with two sets of
Fe\,{\sc i} lines~: the first made of 20 lines with $W \leq 70$ m\AA,
the second made of these plus 11 lines with $70 < W \leq 163$ m\AA\
(see Table~\ref{tab:ato}).
The two studies give very similar results, with negligible differences
($\Delta T_{\rm eff} = 0$, $\Delta \log g = +0.03$,
$\Delta \xi = +0.1$ km s$^{-1}$, $\Delta {\rm
  [Fe/H]} = +0.01$ and a few hundredths of dex for the other elements;
the largest difference being ${\rm [Ca/Fe]} = -0.05$ dex). The
results presented in Table \ref{tab:hd119285} have been obtained with
the large set of lines: ``weak'' plus ``strong''.

As many damping constants are poorly known, all the lines that may have
been significantly van der Walls broadened ($\log (W/\lambda) \geq -4.55$)
were discarded from the analysis. In the remaining transitions, the van der Waals
damping process is a minor contributor to the equivalent widths.
MOOG's Uns\"old approximation option was used to take into account
van der Waals damping in the computation of theoretical curves of growth.

\begin{table*}[t!]
\centering
\caption{Wavelengths, excitation potentials from
\citet{ku95cd23}, $\log gf$ values from this work 
(Ka 02), \citet{ed93} (Ed 93), \citet{ne97} (Ne 97) and \citet{fe01} (Fe 01)
and equivalent widths measured in IS~Vir and V851 Cen. All
$\log gf$ values have been rescaled to Kurucz solar abundances (listed in front
of the element symbols).}
\label{tab:ato}
\begin{tabular}{l c c c c c c c} \hline
                         & $\chi$ &$\log gf$ & $\log gf$ & $\log gf$ & $\log gf$ & IS~Vir & V851~Cen \\
                         & (eV) & (Ka 02)   & (Ed 93)   & (Ne 97)   & (Fe 01)   & W (m\AA)  & W (m\AA)  \\ \hline
{\bf Na I}; $\log Na = 6.33$ & &         &         &          &          &           &           \\
6154.226                 & 2.10 & $-$1.580 & $-$1.61 &          & $-$1.58  & 102.4     & 95.6      \\
{\bf Mg I}; $\log Mg = 7.49$ & &         &         &          &          &           &           \\
5711.088                 & 4.35 & $-$1.653 &         &          &          & 155.1     & 158.4     \\
{\bf Al I}; $\log Al = 6.47$ & &         &         &          &          &           &           \\
6698.673                 & 3.14 & $-$1.867 &         & $-$1.837 & $-$1.89  & 65.7      & 69.5      \\
7084.643                 & 4.02 & $-$1.136 &         &          &          & 57.3      & 66.0      \\
7835.309                 & 4.02 & $-$0.689 &         &          & $-$0.78  & 83.3      & 93.3      \\
{\bf Si I}; $\log Si = 7.55$ & &         &         &          &          &           &           \\
6029.869                 & 5.98 & $-$1.550 &         &          &          & 26.8      & 20.0      \\
6125.021                 & 5.61 & $-$1.490 & $-$1.54 &          & $-$1.55  & 50.8      & 41.8      \\
6155.134                 & 5.62 & $-$0.730 & $-$0.77 &          &          & 82.7      & 72.9      \\
6194.416                 & 5.87 & $-$1.540 &         &          &          & 26.7      & 21.8      \\
6195.433                 & 5.87 & $-$1.570 &         &          &          & 32.3      & 24.8      \\
6237.319                 & 5.61 & $-$1.010 &         &          & $-$1.15  & 74.8      & 67.8      \\
6721.848                 & 5.86 & $-$1.080 &         &          & $-$1.16  & 57.7      &           \\
{\bf Ca I}; $\log Ca = 6.36$ & &          &         &          &          &           &           \\
5867.562                 & 2.93 & $-$1.591 & $-$1.61 & $-$1.534 & $-$1.61  & 76.9      & 68.5      \\
6156.023                 & 2.52 & $-$2.410 &         &          &          & 53.0      & 50.2      \\
6166.439                 & 2.52 & $-$1.100 & $-$1.20 & $-$1.078 & $-$1.22  & 133.4     & 127.4     \\
6455.598                 & 2.52 & $-$1.370 &         & $-$1.344 & $-$1.48  & 125.3     & 120.9     \\
{\bf Sc II}; $\log Sc = 3.10$&          &         &          &          &           &           \\
6245.637                 & 1.51 & $-$1.010 &         & $-$1.066 &          & 81.8      & 78.9      \\
6320.851                 & 1.50 & $-$1.800 &         &          &          &           & 26.9      \\
{\bf Ti I}; $\log Ti = 4.99$ &          &         &          &          &           &           \\
5766.359                 & 3.29 & 0.329    &         &          &          & 46.5      & 46.1      \\
6098.658                 & 3.06 & $-$0.010 &         &          & $-$0.07  & 42.5      & 39.9      \\
{\bf Fe I}; $\log Fe = 7.67$ &          &         &          &          &           &           \\
5833.926                 & 2.61 & $-$3.730 &         &          &          & 65.5      & 47.1      \\
5855.076                 & 4.61 & $-$1.730 & $-$1.72 & $-$1.646 & $-$1.70  & 52.0      & 41.1      \\
5856.088                 & 4.29 & $-$1.750 & $-$1.76 &          &          & 72.0      & 61.1      \\
5861.109                 & 4.28 & $-$2.480 & $-$2.50 & $-$2.491 &          & 34.8      & 24.0      \\
6012.210                 & 2.22 & $-$3.990 &         &          &          & 86.0      & 69.2      \\
6027.051                 & 4.08 & $-$1.290 &         &          &          & 110.2     & 94.0      \\
6056.005                 & 4.73 & $-$0.550 &         &          &          & 106.6     & 92.2      \\
6078.491                 & 4.80 & $-$0.451 &         &          &          & 115.2     & 100.3     \\
6079.008                 & 4.65 & $-$1.150 &         &          &          & 88.5      & 76.3      \\
6093.643                 & 4.61 & $-$1.540 &         & $-$1.443 &          & 58.0      & 45.9      \\
6094.373                 & 4.65 & $-$1.770 &         & $-$1.698 & $-$1.74  & 47.0      & 35.2      \\
6098.244                 & 4.56 & $-$1.970 &         & $-$1.861 &          & 45.5      & 34.2      \\
6151.617                 & 2.18 & $-$3.510 & $-$3.49 & $-$3.404 & $-$3.489 & 122.1     & 103.8     \\
6157.728                 & 4.08 & $-$1.360 & $-$1.40 &          & $-$1.44  & 119.2     & 97.6      \\
6165.360                 & 4.14 & $-$1.670 & $-$1.69 &          & $-$1.684 & 83.8      & 71.9      \\
6307.854                 & 3.64 & $-$3.599 &         &          &          & 25.8      & 17.8      \\
6335.330                 & 2.20 & $-$2.450 &         &          &          &           & 171.9     \\
6336.823                 & 3.69 & $-$0.910 &         &          &          &           & 162.5     \\
6385.718                 & 4.73 & $-$2.010 &         & $-$1.962 & $-$1.98  & 33.8      & 24.0      \\
6436.406                 & 4.19 & $-$2.550 &         &          &          & 39.1      & 29.5      \\
6593.870                 & 2.43 & $-$2.472 &         &          &          &           & 152.2     \\
6699.141                 & 4.59 & $-$2.180 &         & $-$2.215 &          & 27.4      & 19.6      \\
6713.743                 & 4.80 & $-$1.640 &         & $-$1.547 &          & 46.2      & 31.5      \\
6725.356                 & 4.10 & $-$2.380 &         & $-$2.277 &          & 47.3      & 34.3      \\
6726.666                 & 4.61 & $-$1.212 &         &          & $-$1.239 & 78.1      & 65.5      \\
\end{tabular}
\end{table*}
 
\begin{table*}[t!]
\centering
\begin{tabular}{l c c c c c c c} \hline
                         & $\chi$ & $\log gf$ & $\log gf$ & $\log gf$ & $\log gf$ & IS~Vir & V851~Cen \\
                         & (eV) & (Ka 02)   & (Ed 93)   & (Ne 97)   & (Fe 01)   & W (m\AA)  & W (m\AA)  \\ \hline
6750.152                 & 2.42 & $-$2.741 &         &          & $-$2.801 & 159.8     & 134.8     \\
6793.258                 & 4.08 & $-$2.580 &         &          &          & 42.9      & 29.5      \\
6806.843                 & 2.73 & $-$3.300 &         &          & $-$3.27  & 100.4     & 86.2      \\
6810.262                 & 4.61 & $-$1.150 &         &          & $-$1.186 & 87.8      & 76.8      \\
6857.249                 & 4.08 & $-$2.280 &         &          &          & 53.8      & 43.4      \\
6861.937                 & 2.42 & $-$3.990 &         &          &          & 81.7      & 65.1      \\
6862.493                 & 4.56 & $-$1.590 &         &          &          & 62.8      & 50.2      \\
7142.517                 & 4.96 & $-$1.139 &         &          &          & 78.2      & 58.0      \\
7148.725                 & 4.28 & $-$2.297 &         &          &          & 51.6      & 44.2      \\
7411.153                 & 4.28 & $-$0.418 &         &          &          & 157.5     & 132.1     \\
7447.394                 & 4.96 & $-$1.144 &         &          &          & 75.5      & 57.0      \\
7498.530                 & 4.14 & $-$2.320 &         &          & $-$2.29  & 63.1      & 50.5      \\
7563.010                 & 4.84 & $-$1.656 &         &          &          & 44.5      &           \\
7583.788                 & 3.02 & $-$2.040 &         &          &          & 152.4     & 131.8      \\
7748.269                 & 2.95 & $-$1.760 &         &          &          & 184.0     & 165.0     \\
7780.557                 & 4.47 & $-$0.151 &         &          &          &           & 145.5     \\
7807.909                 & 4.99 & $-$0.614 &         &          &          & 90.5      & 81.0      \\
{\bf Fe II}; $\log Fe = 7.67$& &         &         &          &          &           &           \\
5991.376                 & 3.15 & $-$3.727 &         & $-$3.718 &          & 44.6      & 29.9      \\
6084.111                 & 3.20 & $-$3.988 &         & $-$3.968 &          & 31.0      &           \\
6149.258                 & 3.89 & $-$2.874 & $-$2.96 &          & $-$3.00  & 48.1      & 32.1      \\
6247.557                 & 3.89 & $-$2.419 &         &          & $-$2.55  & 63.6      & 48.1      \\
6432.680                 & 2.89 & $-$3.758 &         &          & $-$3.80  & 56.1      & 38.8      \\
6456.383                 & 3.90 & $-$2.235 &         &          & $-$2.40  & 79.4      & 55.5      \\
{\bf Co I}; $\log Co = 4.92$ & &         &         &          &          &           &           \\
6595.864                 & 3.71 & $-$0.647 &         &          &          & 26.7      & 20.8      \\
7838.134                 & 3.97 & $-$0.300 &         &          &          & 25.0      & 21.2      \\
{\bf Ni I}; $\log Ni = 6.25$ & &         &         &          &          &           &           \\
5805.213                 & 4.17 & $-$0.580 &         &          &          & 62.1      & 51.2      \\
6111.066                 & 4.09 & $-$0.790 & $-$0.96 & $-$0.739 & $-$0.86  & 62.2      & 52.2      \\
6176.807                 & 4.09 & $-$0.170 & $-$0.37 &          & $-$0.27  & 90.9      & 80.8      \\
6204.600                 & 4.09 & $-$1.110 & $-$1.17 & $-$1.063 & $-$1.146 & 46.9      & 33.6      \\
6598.593                 & 4.24 & $-$0.860 &         & $-$0.883 & $-$0.94  & 55.4      & 39.9      \\
6772.313                 & 3.66 & $-$0.910 &         &          & $-$1.082 & 83.4      & 70.9      \\
7525.111                 & 3.64 & $-$0.516 &         &          & $-$0.64  & 120.5     & 106.3     \\
7555.598                 & 3.85 & 0.074   &         &          &          & 131.1     & 111.5     \\
7797.586                 & 3.90 & $-$0.172 & $-$0.40 &          &          & 119.7     & 105.2     \\
\end{tabular}
\end{table*}

\subsection{$\log gf$ adjustment} \label{sub:loggf}

Many of the lines used in this analysis don't have accurate oscillator
strengths reported in the literature. We have thus determined, from a
solar spectrum, the oscillator strengths of most of the lines (see
below) used in the
present analysis. Their equivalent widths were measured in a high
$S/N$ ratio ($S/N \simeq 250$) moon spectrum extracted from the
archive of the first FEROS commissioning period (fall 1998). Their
oscillator strengths were then adjusted using a Kurucz
model \rebf{atmosphere with} $T_{\rm eff} = 5777$ K, $\log g = 4.44$ and
micro-turbulence $\xi = 1.0$ km s$^{-1}$, so that a spectral
analysis made with the same approach as for IS Vir and V851 Cen stars
(i.e. deriving abundances with the MOOG LTE analysis software packages)
reproduced the Kurucz solar
abundances.  In order to be consistent with Kurucz models and
opacities, we kept the old Fe\,{\sc i} solar abundance, $\log {\rm Fe}
= 7.67$, instead of the meteoritic value $\log {\rm Fe} = 7.51$.  As
the analysis performed here is purely differential with respect to the
Sun, the use of the ``old'' Fe value has no consequence \rebf{on the}
results derived.  In order to avoid degrading the accuracy of the
oscillator strengths with equivalent widths' measurement errors, only
lines with $W_\odot > 10$ m\AA\ were calibrated on the Sun. For the
weaker lines, we kept the Kurucz theoretical values (\citealp{ku95cd23}).
The characteristics of the selected lines are listed in
Table~\ref{tab:ato}, with their wavelengths, excitation potentials and
adopted oscillator strengths. When available, oscillator strengths
from the literature, extracted from \citet{ed93},
\citet{ne97} and \citet{fe01} are also listed.

\subsection{Continuum fitting and line's measurement} \label{sub:cont}

The FEROS spectra were straightened by segments of 200 to 400 \AA.  In
each of them, intervals of the continuum were selected by comparison
with a Kurucz synthetic spectrum. Those series of intervals were
adjusted by low degree polynomials, which in turn were used to
straighten the different domains. In slow rotators, such as IS~Vir and
V851~Cen, the fitting of the continuum is relatively simple. This is
not true for intermediate rotators \rebf{which} display very few
intervals at continuum level. There is then a risk to underestimate
the position of the continuum. An underestimation of 0.5\% of the
continuum, would lead to an error of 5\% on the equivalent widths and
0.05 dex on the abundances (on ``weak'' lines).

The equivalent widths were measured with the IRAF\footnote{distributed
by the \emph{ National Optical Astronomy Observatories}} 
(\citealp{ja98}) task {\it splot}, assuming Gaussian profiles.
They are compiled in Table~\ref{tab:ato}.

\subsection{Atmospheric models} \label{sub:model}

The analysis were performed \rebf{using} Kurucz' LTE plane parallel atmospheric
models, computed without the overshooting option and for a value of the
length of the convective cell over the pressure scale height
$\alpha = l / H_{\rm p} = 0.5$. This later value has been shown to give
a better description of Balmer $H\alpha$ and $H\beta$ line profiles
than the previous value of 1.25 (\citealp{fu93}, \citealp{va96}).

\begin{table*}[ht!]
\caption{Characteristics of the 9 test stars: galactic coordinates, distances (pc), colour excesses, $B-V$, $V-R_{\rm c}$ and $V-I_{\rm c}$ indices and the
corresponding effective temperatures.}
\label{tab:red}
\begin{center}
\begin{tabular}{l c c c c c c c c c c c c c} \hline
HD Num.   & $l$ & $b$ & $d$  & \scriptsize $E(B-V)$ 
& \scriptsize $E(V-R_{\rm c})$ & \scriptsize $E(V-I_{\rm c})$  & $B-V$ &
$V-R_{\rm c}$ & $V-I_{\rm c}$ & $T_{\rm eff}$ & 
$T_{\rm eff}$ & $T_{\rm eff}$\\ 
 & & & & & & & & & & \scriptsize{(B-V)} & \scriptsize{(V-R)}
& \scriptsize{(V-I)} \\ \hline
 HD 26354 &   262 & $-$46 &   35 & 0.03 & 0.02 & 0.04 &  0.88 & 0.52 & 0.99 &  4963 &  4927 &  4824 \\
 HD 34802 &   289 & $-$32 &  180 & 0.06 & 0.04 & 0.08 &  1.03 & 0.57 & 1.09 &  4746 &  4602 &  4546 \\
 HD 61245 &   258 & $-$12 &  111 & 0.03 & 0.02 & 0.04 &  1.01 & 0.54 & 1.02 &  4785 &  4704 &  4687 \\
 HD 72688 &   255 &     3 &  131 & 0.02 & 0.02 & 0.03 &  0.93 & 0.47 & 0.91 &  4945 &  4973 &  4936 \\
 HD 81410 &   254 &    19 &  120 & 0.04 & 0.03 & 0.06 &  0.98 & 0.53 & 1.02 &  4844 &  4740 &  4687 \\
HD 106225 &   287 &    53 &  125 & 0.01 & 0.00 & 0.01 &  0.99 & 0.58 & 1.13 &  4824 &  4569 &  4472 \\
HD 119285 &   309 &     1 &   76 & 0.03 & 0.02 & 0.05 &  1.05 & 0.60 & 1.15 &  4708 &  4507 &  4436 \\
HD 136905 &   356 &    40 &   95 & 0.07 & 0.04 & 0.09 &  0.96 & 0.53 & 1.01 &  4884 &  4740 &  4708 \\ \hline
\end{tabular}
\end{center}
\end{table*}

\subsection{Effective temperature from photometric indices} \label{sub:photo}

The use of photometric indices to derive effective temperature is a
classical technique commonly used to analyse non-active stars. In the
present work we are however considering stars with a very high level of
\rebf{magnetic} activity, \rebf{which also produces large photospheric
spots}. Their high activity level could thus affect their
spectral energy distribution and therefore their colour indices.  In
the case of the metallicity, large discrepancies between photometric
and spectroscopic estimates were reported by
\citet{gi91} and \citet{fa97}.

In order to check the relative consistency of different colour
indices, we derived, for 8 SB1 objects from our program
(hereafter referred to as test stars), with good quality
Hipparcos parallaxes and $B-V$, $V-R_{\rm c}$ and $V-I_{\rm c}$
indices either from \citet{cu98} or
\citet{cu01}, the effective temperatures corresponding to the three
colours.
\rebf{There was no recent
measurement of the $V-R$ or $V-I$ colour indices reported in the literature
for IS~Vir, which, therefore, has not been included in the test stars sample.}
The visual extinction (i.e. $A(V)$) was
derived using the model of \citet{ar92} and converted in
colour excesses using the coefficients~: $E(B-V) = 0.302\, A(V)$,
$E(V-R_{\rm c}) = 0.194\, A(V)$ and $E(V-I_{\rm c}) = 0.415\, A(V)$
(\citealp{sc98}). Galactic coordinates, distances (pc)
and colour excesses of the tests stars are summarised in Table
\ref{tab:red}.

The colours were converted into effective temperatures
using the empirical calibration for F0V to K5V main sequence stars of
\citet{al96} (HD 26354) and for F0 to K5 giant stars of
\citet{al99} (for the 7 others), initially assuming
solar metallicities. The 8 test stars were classified as either main
sequence or evolved stars, by comparing their absolute magnitudes with
the \citet{be94} isochrones. To convert the $V-R_{\rm c}$
and $V-I_{\rm c}$ indices measured by \citet{cu98} and \citet{cu01} into the
$V-R_{\rm J}$ and $V-I_{\rm J}$ indices used by
\citet{al96} and \citet{al99}, we used the \citet{be79}
transformations, $V-R_{\rm J} = (V-R_{\rm c} + 0.03) / 0.73$ and
$V-I_{\rm J} = V-I_{\rm c} / 0.778$.  The colour indices as well as the
corresponding effective temperatures are summarised in
Table~\ref{tab:red}.

In all cases, the $V-I_{\rm c}$ index lead to
temperatures systematically cooler (from 10 to 350 K with a mean of 175 K)
than the $B-V$ index. In 6 cases out of 8, the $V-R_{\rm c}$ indices give
temperatures significantly cooler than $B-V$ (and slightly warmer
than $V-I_{\rm c}$), with differences ranging
from 80 to 255 K.

In order to check if the discrepancies between the effective temperatures
derived from $B-V$, and those derived from $V-I$, were \aabf{artefacts} of the
adopted transformations, we compared \citet{al99} calibrations
with those for giant stars of \citet{ma90}. In the ranges of
colour relevant to our test stars~: $0.85 \leq B-V \leq 1.10$ and
$0.90 \leq V-I_{\rm c} \leq 1.20$, the differences between the
2 systems never exceed 40 and 20 K. We also computed
the mean differences $T_{\rm eff}(B-V) - T_{\rm eff}(V-I)$ over 7
evolved test stars (i.e. all test stars but HD 26354)
with the 2 calibrations and found
respectively 181 K (\citealp{al99}) and 180~K
(\citealp{ma90}), in perfect agreement.

\citet{al96} and \citet{al99} calibrations have been derived for single
stars and their validity is established only for single systems.
A binary system, in which both components contribute
significantly to the total flux, and at the same time differ in
colours, will deviate from the colour/temperature relations. Of course\rebf{,}
in most cases, a large difference in colour also means a large difference
in brightness and therefore a small contribution of the secondary to
the overall system colours.\\
The possible role of binarity in the discrepancies observed in the
test stars, between the $B-V$
and $V-I$ temperatures, was investigated via a synthetic
grid of binary systems (computed using the isochrones of
\citealp{be94}, for solar metallicity). This grid contains all possible
arrangements of primary and secondary components ranging from the top of the
giant branch to K7 dwarf ($\simeq 3900$ K) by steps of 50 K. Several system
ages have been considered~: 0.5, 1, 2, 3, 4, 5, 8, 10 and 12 Gyr.
Each arrangement is characterised by the effective
temperature\footnote{\rebf{for the global system, two effective
temperatures were derived, one from the $B-V$ index, the other from
the $V-I$ index}}, evolutionary stage (dwarf or giant),
\rebf{visual absolute magnitude and $B-V$ and $V-I$ colour indices of
each component and of the global system}.\\
For each of the 8 test stars, we searched
the grid for the synthetic binary system with similar
$B-V$ temperature and visual absolute magnitude (i.e. within $\pm 1
\sigma$ of the test star parameters) that showed the largest difference
between the temperatures derived from $B-V$ and $V-I$ indices.
To take into account that all the test stars are single line
systems, only systems with secondary component at least 2.5
magnitudes fainter than the primary were considered.\\
In the case of the dwarf star HD 26354, the largest discrepancy between
the $B-V$ and $V-I$ diagnostics is around 150 K \rebf{(assuming a 1 Gyr old
system with a 5200 K subdwarf as primary component and a 3900 K dwarf as
secondary component)}. For the 7 evolved test stars, the
``strongest differences'' range from 18 to 67 K with a mean of 41 K.
This mean value of 41 K is small compared to the mean difference of 181 K
derived from the real colours of the same 7 test stars.
\rebf{Moreover, the above 18 to 67 K differences represent
``extreme'' cases among a
large number of combinations of primary and secondary components, whose
$B-V$ and $V-I$ colour indices lead to very
similar effective temperatures.}
\rebf{Therefore,} the binarity may play a minor role in the
discrepancy between the $B-V$ and $V-I$ temperatures, but is
most likely not the dominant effect.\\

Unlike the $V-I/T_{\rm eff}$ transformation, both the $B-V$ and
$V-R$ vs. temperature relations of \citet{al99} are
sensitive to metallicity. To test whether the
above discrepancy can be solved by adjusting the iron abundances, we
performed a full detailed analysis on 4 of the previous stars,
namely HD 26354, HD 34802, HD 106225 and HD 119285, using the
second method together with the $B-V$ index. The derived metallicities
and corresponding $B-V$ effective temperatures are presented in
Table~\ref{tab:met}. $V-I$ temperatures are also listed for comparison.
The metallicities of HD 34802 and HD 119285
are close to solar and therefore their temperatures are fairly similar
to those listed in Table \ref{tab:red}. HD 26354 exhibits an iron
overabundance which increases the difference between the $B-V$ and
$V-I$ temperatures by about 120 K. The moderate deficiency of HD 106225,
reduces by 100 K the discrepancy between $B-V$ and $V-I$,  but the difference
is still about 250 K. Therefore the metallicity is unlikely to explain the
temperature discrepancy derived through the two indices, discrepancy which
\rebf{may be due to} stellar activity.

\begin{table}[h!]
\caption{Metallicities, $B-V$ and $V-I$ temperatures of 4 of
the test stars. The errors given for $T_{\rm eff}$
correspond to the propagation of the errors on the metallicity.}
\label{tab:met}
\begin{center}
\begin{tabular}{l c c c} \hline
HD Num. & ${\rm [Fe/H]}$ & $T_{\rm eff}$ & $T_{\rm eff}$ \\ 
        &          & \scriptsize $(B-V)$  & \scriptsize $(V-I)$ \\ \hline
 HD 26354 &  $+0.36 \pm 0.22$  &  $5085^{+85}_{-77}$ &  4824 \\
 HD 34802 &  $+0.17 \pm 0.21$  &  $4793^{+65}_{-57}$ &  4546 \\
HD 106225 &  $-0.40 \pm 0.16$  &  $4723^{+38}_{-33}$ &  4472 \\
HD 119285 &  $-0.14 \pm 0.12$  &  $4674^{+29}_{-27}$ &  4436 \\
\end{tabular}
\end{center}
\end{table}

\subsection{Surface gravities by profile fitting} \label{sub:grav}

The wings of ``strong'' lines are \rebf{collisionally} broadened and therefore
gravity sensitive. In the third method, we take advantage of this
behaviour to derive the surface gravity of our stars. The wings of one
of those ``strong'' lines are compared to a library of synthetic lines
spanning a large range in gravities.  In a first step the grid is
searched for the most similar synthetic profile.  In a second step the
difference of gravity between the best synthetic match and the
observed profile is estimated.  Variations of this method have been
successfully used by several authors (e.g.
\citealp{dr91}). Several ``strong'' lines are present in the FEROS
spectral interval: the magnesium green triplet ($\lambda = {5167.3,
  5172.7, 5183.6}$ \AA), the neutral calcium red triplet ($\lambda =
{6102.7, 6122.2, 6162.2}$ \AA) and the ionised calcium infrared triplet
($\lambda = {8498.0, 8542.1, 8662.1}$ \AA). We used the 6162 \AA \,
neutral calcium line whose profile is more sensitive to surface
gravity than the two other transitions of the triplet.  This line
looses relatively quickly its sensitivity to gravity below $\log g =
2$. For more evolved stars it is more accurate to rely on the Ca\,{\sc
 ii} triplet lines, whose strengths increase with decreasing
gravity. A grid of about 1600 synthetic spectra was computed with the
Kurucz' ATLAS9 (\citealp{ku93cd13}) and SYNTHE
(\citealp{ku93cd18}) software packages.  It covers the
parameter space $4250 \leq T_{\rm eff} \leq 6000$ K (by
steps of 250 K), $0.5 \leq \log g \leq 5.0$ (by steps of 0.5), $\xi =
0, 1, 2, 4$ km s$^{-1}$, ${\rm [Fe/H]} = -1, -0.5, -0.3, 0.0, 0.3$ and
$[\alpha/\rm Fe] = 0.0$ and $+0.4$.  Each spectrum ranges from 6000 to
6250 \AA\ with an initial resolution of 300\,000.

The 6162 \AA \, calcium line is not only sensitive to gravity, but
also to effective temperature, metallicity,
micro-turbulence and, of course, calcium abundance. As a
consequence, the first step \rebf{in} the estimation of gravity is to fix
those four parameters.  The synthetic grid is then interpolated to
compute a new sub-grid at the chosen values of $T_{\rm eff}$, $\xi$,
${\rm [Fe/H]}$ and ${\rm [Ca/H]}$, keeping $\log g$ as the only degree
of freedom. The 10 spectra of this sub-grid are then compared \aabf{with} the
observed spectrum. For each comparison, the flux of the synthetic and
observed spectra are adjusted by least squares, finding the
coefficients $a_0$ and $a_1$ that minimise the expression:
\begin{equation}
S = {1 \over n - 2} \sum_{i = 1}^n (F_{\rm obs}(i) - (a_0 + a_1 \lambda(i))
F_{\rm syn}(i))^2 B(i) 
\end{equation}

where $n$ is the number of pixels compared between the two spectra,
$F_{\rm obs}$ and $F_{\rm syn}$ are the respective flux values in the
observed and synthetic spectra, $\lambda(i)$ is the
wavelength\footnote{the observed spectrum wavelength are corrected
for the radial velocity shift.} and $B$ the profile of the blaze
function of the observed spectrum. Two degrees of freedom are used in
the flux adjustment, in order to compensate a possible relative slope
between the observed and synthetic spectra. A mask is used to compare
only precise portions of the spectra around the 6162 \AA \, line: one
interval on each side of the line to set the level of the continuum
(6158.45--6158.95 \AA, 6164.00--6164.75 \AA) and one interval in each wing
of the line (6161.55--6162.02 \AA, 6162.37--6163.00 \AA).  We avoid using
the central part of the line, formed in the higher layer of the
atmosphere, partially in NLTE conditions and which is less sensitive to
surface gravity, as well as possibly affected by activity. The degree of
similarity between each synthetic spectrum and the observed one is
quantified by the residuals of the fluxes adjustment $S$.

\begin{figure}[h!]
\resizebox{\hsize}{!}
{\rotatebox{90}{\includegraphics{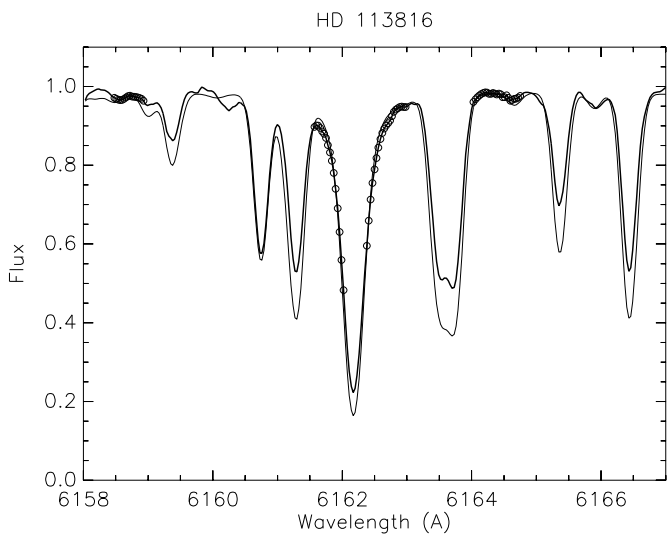}}}
\caption{Comparison of a portion the IS~Vir spectrum (thick line) and the
  interpolated profile (thin line). The open circles show the pixels
  that have been used in the interpolation.}
\label{fig:grav}
\end{figure}

Once the most similar synthetic spectrum (hereafter referred as pivot)
is identified, the difference of gravity between this ``twin''
synthetic profile and the observed one is derived, using a variant of
the atmospheric parameters optimal extraction method
(\citealp{ca91}). The 6162 \AA\ Ca\,{\sc i} profile is projected on the
grid, assuming that between two neighbouring synthetic spectra, the
gravity evolved linearly with the logarithm of the flux: $\log F_{\rm
  obs} = (1-u) \log F_{\rm piv} + u \log F_{\rm nei}$, where $F_{\rm
  obs}$, $F_{\rm piv}$ and $F_{\rm nei}$ are respectively the observed
and pivot fluxes and the flux of the one out of the pivot's two
synthetic neighbours that is most similar to the observed profile.
The projection is performed by least squares, searching for the
coefficients $u$, $a_0$ and $a_1$ that minimise the expression:
\begin{equation}
S' = \sum_{i = 1}^n (\log(F_{\rm obs}/F_{\rm piv}) - u \log(F_{\rm
  nei}/F_{\rm piv}) - a_0 - a_1 \lambda(i))^2
\end{equation}
where $a_0$ and $a_1$ account for a possible difference between the
slopes of the observed and synthetic spectra.

The gravity of the star is then given by
\begin{equation}
\log g = \log g_{\rm piv} + u (\log g_{\rm nei} - \log g_{\rm piv})
\end{equation}

where $\log g_{\rm piv}$ is the surface gravity of the pivot and $\log
g_{\rm nei}$ the gravity of the best synthetic neighbour.

\begin{table}[h!]
  \caption{Parameters of the reference stars used to estimate the
    accuracy of calcium line fitting technique~:
    signal to noise ratios (S/N), effective temperatures ($T_{\rm eff}$),
    surface gravities ($(\log g)_{bib}$), metallicities (${\rm [Fe/H]}$)
    and micro-turbulences ($\xi$). ($\log g$) gives the
    gravities derived by profile fitting.}
\label{tab:grav}
\begin{tabular}{l r c c r c c} \hline
Id. & $S/N$ & $T_{\rm eff}$ & $(\log g)_{bib}$ & ${\rm [Fe/H]}$ & $\xi$ &
$\log g$ \\ \hline 
Sun       & 380 & 5777 & 4.44 &  0.00 $\ $   & 1.0 & 4.42 \\
Sun       & 270 &  ``  &  ``  &  `` $\ \ \ $ & ``  & 4.39 \\
Sun       & 140 &  ``  &  ``  &  `` $\ \ \ $ & ``  & 4.43 \\
Sun       & 160 &  ``  &  ``  &  `` $\ \ \ $ & ``  & 4.43 \\
Sun       & 120 &  ``  &  ``  &  `` $\ \ \ $ & ``  & 4.41 \\
Sun       & 200 &  ``  &  ``  &  `` $\ \ \ $ & ``  & 4.48 \\
Sun       & 220 &  ``  &  ``  &  `` $\ \ \ $ & ``  & 4.46 \\
HD 1835   & 140 & 5771 & 4.44 & 0.15 $\ $    & 1.0 & 4.56 \\
HD 10307  & 200 & 5882 & 4.33 & 0.02 $\ $    & 1.0 & 4.24 \\
HD 24040  & 150 & 5594 & 4.50 & 0.07 $\ $    & 1.0 & 3.95 \\
HD 28099  & 110 & 5761 & 4.50 & 0.17 $\ $    & 1.0 & 4.48 \\
HD 85503  & 180 & 4540 & 2.20 & 0.29 $\ $    & 1.2 & 2.14 \\
HD 86728  & 130 & 5742 & 4.21 & 0.12 $\ $    & 1.0 & 4.37 \\
HD 95128  & 180 & 5855 & 4.25 & 0.00 $\ $    & 1.0 & 4.14 \\
HD 113226 & 140 & 4990 & 2.70 & 0.11 $\ $    & 1.6 & 2.33 \\
HD 114710 &  90 & 5979 & 4.40 & 0.08 $\ $    & 1.0 & 4.22 \\
HD 115383 & 110 & 5920 & 3.96 & 0.10 $\ $    & 1.3 & 4.16 \\
HD 146233 & 240 & 5803 & 4.34 & 0.03 $\ $    & 1.0 & 4.48 \\
HD 184406 &  80 & 4450 & 2.47 & $-0.13\ \ $  & 1.8 & 2.64 \\
HD 186408 & 160 & 5820 & 4.26 & 0.07 $\ $    & 1.0 & 4.31 \\
HD 186427 & 140 & 5762 & 4.38 & 0.06 $\ $    & 1.0 & 4.39 \\
HD 187691 & 160 & 6101 & 4.22 & 0.09 $\ $    & 1.0 & 4.21 \\
HD 217014 & 160 & 5757 & 4.23 & 0.15 $\ $    & 1.0 & 4.24 \\
HD 219134 & 150 & 4727 & 4.50 & 0.05 $\ $    & 1.0 & 4.55
\end{tabular}
\end{table}

\begin{table*}[t!]
\caption{IS Vir: number of transitions used to derive the abundances of the
different elements (n), mean values ($<>$) and error bars ($\sigma$) of the
atmospheric parameters and abundances, with the three methods.}
\label{tab:hd113816}
\begin{center}
\begin{tabular}{c c c c c c c c c c c c}
                  & \multicolumn{3}{c}{meth. 1} &
                  & \multicolumn{3}{c}{meth. 2} &
                  & \multicolumn{3}{c}{meth. 3} \\ \hline
                  & n & $<>$ & $\sigma$ & \, \, \, \, \,
                  & n & $<>$ & $\sigma$ & \, \, \, \, \,
                  & n & $<>$ & $\sigma$ \\ \hline
$T_{\rm eff}$     & & 4720      & 100 &
                  & & 4690      & 150 &
                  & & 4780      & 185 \\
$\log g$          & & 2.65      & 0.25 &
                  & & 2.55      & 0.41 &
                  & & 2.75      & 0.45 \\
$\xi$             & & 1.65      & 0.08 &
                  & & 1.65      & 0.11 &
                  & & 1.60      & 0.13 \\
${\rm [Fe/H]}$  & 44 & $+0.04$ & 0.08 & & 36 & $+0.02$ & 0.13 &
                  & 36 & $+0.09$ & 0.16 \\
${\rm [Na/Fe]}$ &  1 & $+0.23$ & 0.04 & & 1 & $+0.23$ & 0.04 &
                  &  1 & $+0.23$ & 0.04 \\
${\rm [Mg/Fe]}$ &  1 & $+0.17$ & 0.06 & & 1 & $+0.18$ & 0.06 &
                  &  1 & $+0.18$ & 0.06 \\
${\rm [Al/Fe]}$ &  3 & $+0.18$ & 0.06 & & 3 & $+0.19$ & 0.07 &
                  &  3 & $+0.16$ & 0.08 \\
${\rm [Si/Fe]}$ &  7 & $+0.06$ & 0.05 & & 7 & $+0.07$ & 0.08 &
                  &  7 & $+0.02$ & 0.11 \\ 
${\rm [Ca/Fe]}$ &  4 & $+0.11$ & 0.04 & & 4 & $+0.10$ & 0.04 &
                  &  4 & $+0.13$ & 0.06 \\
${\rm [Sc/Fe]}$ &  1 & $+0.05$ & 0.05 & & 1 & $+0.03$ & 0.09 &
                  &  1 & $+0.08$ & 0.12 \\
${\rm [Ti/Fe]}$ &  2 & $+0.05$ &  0.04 & & 2 & $+0.04$ & 0.04 &
                  &  2 & $+0.06$ & 0.05 \\
${\rm [Co/Fe]}$ &  2 & $-0.04$ &  0.07 & & 2 & $-0.04$ & 0.07 &
                  &  2 & $-0.04$ & 0.08 \\
${\rm [Ni/Fe]}$ &  9 & $-0.10$ & 0.03 & & 9 & $-0.10$ & 0.04 &
                  &  9 & $-0.10$ & 0.04 \\
\end{tabular}
\end{center}
\end{table*}

\begin{table*}[t!]
\caption{V851 cen: number of transitions used to derive the abundances of the
different
elements (n), mean values ($<>$) and error bars ($\sigma$) of the
atmospheric parameters and abundances, with the three methods.}
\label{tab:hd119285}
\begin{center}
\begin{tabular}{c c c c c c c c c c c c c c c c}
                  & \multicolumn{3}{c}{meth. 1}  &
                  & \multicolumn{7}{c}{meth. 2 (B-V/V-I)} &
                  & \multicolumn{3}{c}{meth. 3} \\ \hline
                  & n & $<>$ & $\sigma$ & \, \, \, \,
                  & n & $<>$ & $\sigma$ & \, \, \, \,
                  & n & $<>$ & $\sigma$ & \, \, \, \,
                  & n & $<>$ & $\sigma$ \\ \hline
$T_{\rm eff}$ &   &   4700      & 80 &
                  & & 4670      & 150 &
                  & & 4440      & 100 &
                  & & 4780      & 160 \\
$\log g$          & & 3.00      & 0.27 &
                  & & 2.90      & 0.42 &
                  & & 2.25      & 0.32 &
                  & & 3.23      & 0.41 \\
$\xi$             & & 1.50      & 0.07 &
                  & & 1.50      & 0.12 &
                  & & 1.60      & 0.10 &
                  & & 1.40      & 0.13 \\
${\rm [Fe/H]}$    & 46 & $-0.13$ & 0.07 & & 36 & $-0.14$ & 0.12 &
                  & 36 & $-0.31$ & 0.10 & & 36 & $-0.04$ & 0.15 \\
${\rm [Na/Fe]}$   &  1 & $+0.30$ & 0.03 & &  1 & $+0.29$ & 0.04 &
                  &  1 & $+0.28$ & 0.04 & &  1 & $+0.27$ & 0.04 \\
${\rm [Mg/Fe]}$   &  1 & $+0.41$ & 0.07 & &  1 & $+0.39$ & 0.07  &
                  &  1 & $+0.44$ & 0.07 & &  1 & $+0.35$ & 0.04 \\
${\rm [Al/Fe]}$   &  3 & $+0.46$ & 0.07 & &  3 & $+0.46$ & 0.08 &
                  &  3 & $+0.51$ & 0.10 & &  3 & $+0.41$ & 0.08 \\
${\rm [Si/Fe]}$   &  6 & $+0.15$ & 0.04 & &  6 & $+0.14$ & 0.07 &
                  &  6 & $+0.26$ & 0.06 & &  6 & $+0.08$ & 0.09 \\ 
${\rm [Ca/Fe]}$   &  4 & $+0.22$ & 0.04 & &  4 & $+0.21$ & 0.06 &
                  &  4 & $+0.14$ & 0.05 & &  4 & $+0.22$ & 0.06 \\
${\rm [Sc/Fe]}$   &  2 & $+0.24$ & 0.10 & &  2 & $+0.22$ & 0.13 &
                  &  2 & $+0.06$ & 0.10 & &  2 & $+0.30$ & 0.15 \\
${\rm [Ti/Fe]}$   &  2 & $+0.19$ & 0.04 & &  2 & $+0.17$ & 0.05 &
                  &  2 & $+0.12$ & 0.05 & &  2 & $+0.18$ & 0.05 \\
${\rm [Co/Fe]}$   &  2 & $+0.07$ & 0.04 & &  2 & $+0.06$ & 0.04 &
                  &  2 & $+0.04$ & 0.04 & &  2 & $+0.06$ & 0.05 \\
${\rm [Ni/Fe]}$   &  9 & $-0.06$ & 0.03 & &  9 & $-0.07$ & 0.04 &
                  &  9 & $-0.09$ & 0.03 & &  9 & $-0.06$ & 0.04 \\
\end{tabular}
\end{center}
\end{table*}

The reliability of the method was tested on a subsample of the
ELODIE\footnote{in operation on the 193 cm telescope from the
  Haute-Provence observatory.} library of standard stars
(\citealp{so98}; \citealp{pr01}). This library, in
addition to the spectra, provides values of $T_{\rm eff}$, $\log g$
and ${\rm [Fe/H]}$ for each star. The ELODIE spectra are relatively
similar in characteristics to FEROS spectra, with a resolution of
42\,000 and a wavelength domain ranging from 3900 to 6800~\AA. We
applied the gravity determination method to 24 spectra from the
library: 7 spectra of the Sun (Moon), 17 main sequence star spectra
and 4 evolved star spectra. We assumed a calcium abundance proportional to
the solar one (${\rm [Ca/Fe]} = 0.0$) for all stars, a
micro-turbulence of 1 km s$^{-1}$ for the main sequence stars and
searched the literature for the micro turbulence of the evolved
objects. Table \ref{tab:grav} list the stars used, their adopted
atmospheric parameters and the surface gravities estimated with our
method. The average difference between the estimated and the
bibliographic gravities is $-0.02$ with a dispersion of 0.16. 
A test performed on IS~Vir has shown that systematically shifting
the calcium line fitting gravity diagnostic by
$\Delta \log g = -0.16$ modifies the third method result
by $\Delta T_{\rm eff} = -140$ K, $\Delta \log g = -0.38$,
$\Delta {\rm [Fe/H]} = -0.13$ dex and $\Delta \xi = +0.1$ km s$^{-1}$.
This error is accounted for in the error bars given for IS~Vir and
V851~Cen in Tables \ref{tab:hd113816} and \ref{tab:hd119285}.

\section{Results} \label{sub:res}

The atmospheric parameters and abundances derived for IS~Vir and V851~Cen,
with the three methods \rebf{outlined in Sect. \ref{sec:data}},
are presented in
Tables \ref{tab:hd113816} and \ref{tab:hd119285}. 
The error bars were derived in several steps. First, the 1~$\sigma$ errors
on each of the ``individual'' diagnostics (e.g. average slope of Fe {\sc i}
abundances versus excitation potentials relation)
were calculated. They were then propagated to obtain the ``individual''
errors on the atmospheric parameters and abundances. Finally, those
``individual'' errors were quadratically summed.

Our \rebf{concern} with the first method was that Fe {\sc i}
``low excitation potential''
transitions may be affected by non-LTE effects. This does not seem
the case neither \rebf{for} IS~Vir nor \rebf{for}
V851~Cen since, in both cases, methods 1 and 3 (the first using 
Fe {\sc i} ``low excitation potential'' iron transitions and the other not)
give results in good agreement.  \rebf{Cooler} or more evolved stars might,
of course, exhibit stronger departure from LTE and would require some
careful checking.

The $B-V$ index leads to \rebf{effective} temperatures in very good agreement
with those derived with the first method and 90 and 110 K \rebf{lower,
for IS~Vir and V851~Cen respectively, than those derived with the third
method. Given the error bars of the second and third methods (respectively
150 and 185 K for IS~Vir and 150 and 160 K for V851~Cen), their results
are compatible within less than one sigma error.} As already presented
in Sect.~\ref{sub:photo}, the temperature derived for V851~Cen from the
$V-I_{\rm c}$ index is about 230 K lower than the one derived from $B-V$.
This is respectively 260 and 340 K lower than the temperatures obtained
via the first and third methods.
Neither IS~Vir, nor V851~Cen display a significant departure
from the $B-V$/temperature relation of non-active stars,
whereas the $V-I$ index deviate from the $V-I$/temperature
calibration. \rebf{This behaviour may be due to the presence of
activity.}

The third method exhibits fairly large error bars. This is in part due
to its dependency on the knowledge of the calcium abundance, which
relies on four lines (2 of them being on the ``plateau'' of the curve
of growth) and in part due to the propagation, \rebf{on the final results,
of the error bars on the surface gravity (derived by calcium wings fitting,
see Sect. \ref{sub:grav}).} Nonetheless it has the merit of
relying on criteria which
present few risks of being affected by non-LTE or activity effects,
as well as being self-consistent. As discussed above, the
results of methods 1 and 3 are consistent.

\section{Discussion} \label{sec:dis}

The following discussion is based on the parameters derived with the
first method (using both ``low'' and ``high'' excitation potential
transitions). Opting for the parameters derived with one of the two
other methods (at the exception of $V-I$ temperature diagnostic)
would not significantly alter the conclusions.

\begin{figure}[t!]
\resizebox{\hsize}{!}
{\rotatebox{90}{\includegraphics{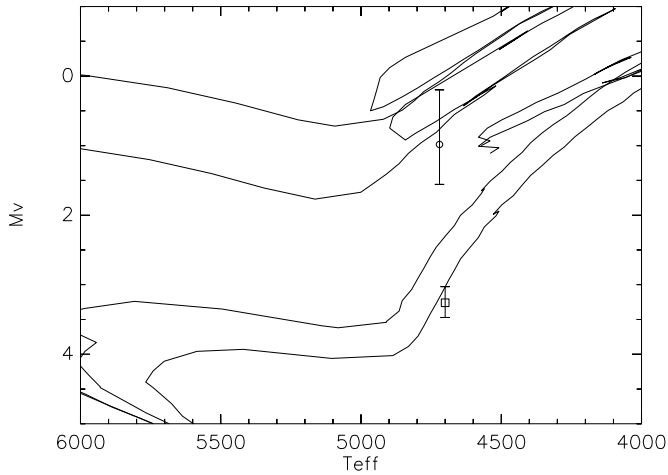}}}
\caption{The position of IS~Vir (circle) and V851~Cen (square) in the
  H-R diagram using the temperatures derived here, plotted with four
  solar metallicity isochrones of respective ages 0.5, 1.0, 5.0 and
  10.0 Gyr (from top to bottom) from \citet{be94}.}
\label{fig:iso}
\end{figure}

{\bf IS~Vir} is a giant star of near-solar metallicity. Its position
in the H-R diagram \rebf{(using $T_{\rm eff}$ determined in this work
and $M_v$ derived from
Hipparcos parallaxes)} is plotted \rebf{on} Fig. \ref{fig:iso} (circle),
together with four isochrones of respective ages
0.5, 1.0, 5.0 and 10 Gyr, \rebf{derived using solar metallicity}
(\citealp{be94}). The mass and bolometric magnitude tabulated by
\citet{be94} for a 1 Gyr old star with $T_{\rm eff} = 4730$ K,
$M = 2.06\, M_\odot$ and $M_{\rm bol} = 0.42$, correspond
to a surface gravity of $\log g = 2.67$, in perfect \rebf{agreement}
with the spectroscopic gravity \rebf{derived here}. This area of
the H-R diagram is too dense to
determine the evolutionary stage of the star, i.e. whether it is at
the beginning of the giant branch or a giant of the clump, or its age
(error bars on $T_{\rm eff}$ are still compatible at the 1.5 $\sigma$
level with a 5 or 10 Gyr clump star\footnote{the gravities \rebf{derived
from the isochrones,} for a 4570 K clump star, are $\log g = 2.4$ and
2.3 respectively for 5 and 10 Gyr.}).  Its chemical mix is somewhat
different from the the solar one, with Na, Mg, Al and Ca exhibiting
overabundances of 0.1 to 0.2 dex with respect to Fe. The Na and Mg
abundances should be considered with caution, since they both rely on
a single line of the curve of growth ``plateau''. The ratio [Ni/Fe] is
somewhat lower than in the Sun.

{\bf V851~Cen} is a slightly metal-poor evolved star. Its position in
the H-R diagram (Fig. \ref{fig:iso}, square) shows \aabf{that it is}
a several Gyr old
star, starting the ascension of the giant branch. The 10 Gyr isochrone
leads to a gravity of $\log g = 3.25$ for a temperature of 4700 K, in
reasonable agreement with the spectroscopic determination. Its
chemical composition displays some similarities with the one of IS~Vir,
with (stronger) overabundances of Na, Mg, Al and Ca ranging from
$+0.2$ to $+0.5$ dex. In addition Si, Sc and Ti are also over-abundant by
$+0.15$ to $+0.25$ dex.

V851~Cen is one of the stars studied by \citet{ra93}.
With a value of $B-V = 1.07$ they derived an effective temperature of
$T_{\rm eff} = 4650$ K, very close to our ($B - V$) photometric determination
($T_{\rm eff} = 4670$ K) and 50 K colder than our first method
estimate. They found a gravity larger by $\Delta \log g =
0.60$.  The largest discrepancy between our analysis and theirs
concern the metallicity, as they derived an iron abundance of ${\rm
  [Fe/H]} = -0.6$, $\simeq -0.5$ dex lower than our determination.

As a caveat, it should be reminded that abundances have been derived,
here, under the assumption of an atmosphere which can be represented
through a single plane parallel model and that the lines were formed
under LTE conditions. Clearly, the behaviour of the colour indices
described above shows that this assumption is likely to be a
simplification of reality in these stars. On the other hand, it is
reassuring to note that three different methods lead to converging
results. However, as shown in Tables~\ref{tab:hd113816} and
\ref{tab:hd119285}\rebf{,} while the absolute abundance values depend on the
effective temperature, most abundance ratios (i.e. Na, Mg, Al, Ca, Ti,
Co and Ni over Fe) appears relatively insensitive to the assumed
temperature.  Therefore, even in an atmosphere which is a
superposition of models with different effective temperatures\rebf{,} the
abundance pattern should remain largely unchanged. In general, all the
other analyses of abundances in active stars have also been carried
out under the same assumptions, so that the resulting bias, if
present, should be common to all works.

\section{Conclusions} \label{sec:concl}

The high-resolution optical spectra of the two X-ray active binaries
RS CVn stars HD 113816 (IS Vir) and HD 119285 (V851 Cen) have been
analysed, using three different techniques. The techniques have been
compared in order to establish their respective merits and drawbacks.
The present results show that the Fe\,~{\sc i} low excitation
potential transitions do not appear significantly affected by non-LTE
effects, and\rebf{,} as a consequence, the iron excitation equilibrium is
a reliable diagnostic in the two studied RS CVn. At the same time,
the $V-I$ photometric index gives significantly colder
(and by comparison with the other approaches, likely underestimated)
effective temperatures than the $B-V$ index.

Both IS~Vir and V851~Cen are found to be near-solar metallicity
giants, with the latter somewhat underabundant in Fe (but
significantly more metal rich, for V851~Cen, than found by
\citealp{ra94}).  Both stars present overabundances with respect to
the solar pattern of their Na, Mg, Al and Ca to Fe ratios. In addition
IS~Vir exhibits a modest underabundance in Ni, while V851~Cen shows
overabundances of Si, Sc and Ti. The analysis of a large sample will
be interesting to see whether this apparent peculiar abundance pattern
is a general characteristic of active binaries. 

\begin{acknowledgements}
  We are very grateful to R. Kurucz, C. Sneden and the VALD people for
  making their software packages and atomic and molecular data
  available to the community. We would like to thank F. Arenou, B.
  Barbuy, V. Hill, M. Spite, F. Spite, C.~van~'t~Veer, I. Pillitteri
  and J. Drake for the very fruitful discussions and advice.

\end{acknowledgements}

\bibliographystyle{apj}
\bibliography{ref}

\begin{thebibliography}{38}
\expandafter\ifx\csname natexlab\endcsname\relax\def\natexlab#1{#1}\fi

\bibitem[{{Alonso} {et~al.}(1996){Alonso}, {Arribas}, \& {Mart{\'
  i}nez-Roger}}]{al96}
{Alonso}, A., {Arribas}, S., \& {Mart{\' i}nez-Roger}, C. 1996, A\&A, 313, 873

\bibitem[{{Alonso} {et~al.}(1999){Alonso}, {Arribas}, \& {Mart{\'
  i}nez-Roger}}]{al99}
---. 1999, A\&AS, 140, 261

\bibitem[{{Arenou} {et~al.}(1992){Arenou}, {Grenon}, \& {Gomez}}]{ar92}
{Arenou}, F., {Grenon}, M., \& {Gomez}, A. 1992, A\&A, 258, 104

\bibitem[{{Bertelli} {et~al.}(1994){Bertelli}, {Bressan}, {Chiosi}, {Fagotto},
  \& {Nasi}}]{be94}
{Bertelli}, G., {Bressan}, A., {Chiosi}, C., {Fagotto}, F., \& {Nasi}, E. 1994,
  A\&AS, 106, 275

\bibitem[{{Bessell}(1979)}]{be79}
{Bessell}, M.~S. 1979, PASP, 91, 589

\bibitem[{{Cayrel} {et~al.}(1991){Cayrel}, {Perrin}, {Barbuy}, \&
  {Buser}}]{ca91}
{Cayrel}, R., {Perrin}, M.-N., {Barbuy}, B., \& {Buser}, R. 1991, A\&A, 247,
  108

\bibitem[{{Cowley} \& {Castelli}(2002)}]{co02}
{Cowley}, C.~R. \& {Castelli}, F. 2002, A\&A, 387, 595

\bibitem[{{Cutispoto}(1998)}]{cu98}
{Cutispoto}, G. 1998, A\&AS, 131, 321

\bibitem[{{Cutispoto} {et~al.}(2001){Cutispoto}, {Messina}, \& {Rodon{\`
  o}}}]{cu01}
{Cutispoto}, G., {Messina}, S., \& {Rodon{\` o}}, M. 2001, A\&A, 367, 910

\bibitem[{{Drake} \& {Smith}(1991)}]{dr91}
{Drake}, J.~J. \& {Smith}, G. 1991, MNRAS, 250, 89

\bibitem[{{Edvardsson} {et~al.}(1993){Edvardsson}, {Andersen}, {Gustafsson},
  {Lambert}, {Nissen}, \& {Tomkin}}]{ed93}
{Edvardsson}, B., {Andersen}, J., {Gustafsson}, B., {Lambert}, D.~L., {Nissen},
  P.~E., \& {Tomkin}, J. 1993, A\&A, 275, 101

\bibitem[{{Favata} {et~al.}(1997){Favata}, {Micela}, {Sciortino}, \&
  {Morale}}]{fa97}
{Favata}, F., {Micela}, G., {Sciortino}, S., \& {Morale}, F. 1997, A\&A, 324,
  998

\bibitem[{{Feldman}(1998)}]{fe98}
{Feldman}, U. 1998, Space Science Reviews, 85, 227

\bibitem[{{Feltzing} \& {Gonzalez}(2001)}]{fe01}
{Feltzing}, S. \& {Gonzalez}, G. 2001, A\&A, 367, 253

\bibitem[{{Fuhrmann} {et~al.}(1993){Fuhrmann}, {Axer}, \& {Gehren}}]{fu93}
{Fuhrmann}, K., {Axer}, M., \& {Gehren}, T. 1993, A\&A, 271, 451

\bibitem[{{Gimenez} {et~al.}(1991){Gimenez}, {Reglero}, {de Castro}, \&
  {Fernandez-Figueroa}}]{gi91}
{Gimenez}, A., {Reglero}, V., {de Castro}, E., \& {Fernandez-Figueroa}, M.~J.
  1991, A\&A, 248, 563

\bibitem[{{Jacoby}(1998)}]{ja98}
{Jacoby}, G. 1998, American Astronomical Society Meeting, 30, 906

\bibitem[{{Jordan} {et~al.}(1998){Jordan}, {Doschek}, {Drake}, {Galvin}, \&
  {Raymond}}]{jo98}
{Jordan}, C., {Doschek}, G.~A., {Drake}, J.~J., {Galvin}, A.~B., \& {Raymond},
  J.~C. 1998, in ASP Conf. Ser. 154: Cool Stars, Stellar Systems, and the Sun,
  Vol.~10, 91

\bibitem[{{Kupka} {et~al.}(1999){Kupka}, {Piskunov}, {Ryabchikova}, {Stempels},
  \& {Weiss}}]{ku99}
{Kupka}, F., {Piskunov}, N., {Ryabchikova}, T.~A., {Stempels}, H.~C., \&
  {Weiss}, W.~W. 1999, A\&AS, 138, 119

\bibitem[{{Kupka} {et~al.}(2000){Kupka}, {Ryabchikova}, {Piskunov}, {Stempels},
  \& {Weiss}}]{ku00}
{Kupka}, F.~G., {Ryabchikova}, T.~A., {Piskunov}, N.~E., {Stempels}, H.~C., \&
  {Weiss}, W.~W. 2000, Baltic Astronomy, 9, 590

\bibitem[{{Kurucz}(1993{\natexlab{a}})}]{ku93cd13}
{Kurucz}, R. 1993{\natexlab{a}}, ATLAS9 Stellar Atmosphere Programs and 2 km/s
  grid.~Kurucz CD-ROM No.~13.~ Cambridge, Mass.: Smithsonian Astrophysical
  Observatory, 1993., 13

\bibitem[{{Kurucz}(1993{\natexlab{b}})}]{ku93cd18}
---. 1993{\natexlab{b}}, SYNTHE Spectrum Synthesis Programs and Line
  Data.~Kurucz CD-ROM No.~18.~Cambridge, Mass.: Smithsonian Astrophysical
  Observatory, 1993., 18

\bibitem[{{Kurucz} \& {Bell}(1995)}]{ku95cd23}
{Kurucz}, R. \& {Bell}, B. 1995, Atomic Line Data (R.L.~Kurucz and B.~Bell)
  Kurucz CD-ROM No.~23.~Cambridge, Mass.: Smithsonian Astrophysical
  Observatory, 1995., 23

\bibitem[{{McWilliam}(1990)}]{ma90}
{McWilliam}, A. 1990, ApJS, 74, 1075

\bibitem[{{Moore} {et~al.}(1966){Moore}, {Minnaert}, \& {Houtgast}}]{mo66}
{Moore}, C.~E., {Minnaert}, M.~G.~J., \& {Houtgast}, J. 1966, {The solar
  spectrum 2935 A to 8770 A} (National Bureau of Standards Monograph,
  Washington: US Government Printing Office (USGPO), 1966)

\bibitem[{{Neuforge-Verheecke} \& {Magain}(1997)}]{ne97}
{Neuforge-Verheecke}, C. \& {Magain}, P. 1997, A\&A, 328, 261

\bibitem[{{Ottmann} {et~al.}(1998){Ottmann}, {Pfeiffer}, \& {Gehren}}]{ot98}
{Ottmann}, R., {Pfeiffer}, M.~J., \& {Gehren}, T. 1998, A\&A, 338, 661

\bibitem[{{Piskunov} {et~al.}(1995){Piskunov}, {Kupka}, {Ryabchikova}, {Weiss},
  \& {Jeffery}}]{pi95}
{Piskunov}, N.~E., {Kupka}, F., {Ryabchikova}, T.~A., {Weiss}, W.~W., \&
  {Jeffery}, C.~S. 1995, A\&AS, 112, 525

\bibitem[{{Prugniel} \& {Soubiran}(2001)}]{pr01}
{Prugniel}, P. \& {Soubiran}, C. 2001, A\&A, 369, 1048

\bibitem[{{Randich} {et~al.}(1994){Randich}, {Giampapa}, \&
  {Pallavicini}}]{ra94}
{Randich}, S., {Giampapa}, M.~S., \& {Pallavicini}, R. 1994, A\&A, 283, 893

\bibitem[{{Randich} {et~al.}(1993){Randich}, {Gratton}, \&
  {Pallavicini}}]{ra93}
{Randich}, S., {Gratton}, R., \& {Pallavicini}, R. 1993, A\&A, 273, 194

\bibitem[{{Rice} \& {Strassmeier}(2001)}]{ri01}
{Rice}, J.~B. \& {Strassmeier}, K.~G. 2001, A\&A, 377, 264

\bibitem[{{Ruland} {et~al.}(1980){Ruland}, {Biehl}, {Holweger}, {Griffin}, \&
  {Griffin}}]{ru80}
{Ruland}, F., {Biehl}, D., {Holweger}, H., {Griffin}, R., \& {Griffin}, R.
  1980, A\&A, 92, 70

\bibitem[{{Schlegel} {et~al.}(1998){Schlegel}, {Finkbeiner}, \& {Davis}}]{sc98}
{Schlegel}, D.~J., {Finkbeiner}, D.~P., \& {Davis}, M. 1998, ApJ, 500, 525

\bibitem[{{Sneden}(1973)}]{sn73}
{Sneden}, C.~A. 1973, Ph.D.~Thesis University of Texas, Austin

\bibitem[{{Soubiran} {et~al.}(1998){Soubiran}, {Katz}, \& {Cayrel}}]{so98}
{Soubiran}, C., {Katz}, D., \& {Cayrel}, R. 1998, A\&AS, 133, 221

\bibitem[{{Strassmeier} {et~al.}(1993){Strassmeier}, {Hall}, {Fekel}, \&
  {Scheck}}]{st93}
{Strassmeier}, K.~G., {Hall}, D.~S., {Fekel}, F.~C., \& {Scheck}, M. 1993,
  A\&AS, 100, 173

\bibitem[{{van't Veer-Menneret} \& {Megessier}(1996)}]{va96}
{van't Veer-Menneret}, C. \& {Megessier}, C. 1996, A\&A, 309, 879

\end{thebibliography}

\end{document}